\documentclass[aps,prd,twocolumn,showkeys,floats,showpacs, superscriptaddress,nofootinbib]{revtex4-1}

\usepackage{graphicx}

\usepackage{bm}
\usepackage{amsmath,amssymb}
\usepackage{aas_macros}
\usepackage{natbib}
\usepackage{textcomp}
\usepackage{multirow}
\usepackage{placeins}
\usepackage[colorlinks=true]{hyperref}

\hypersetup{colorlinks,citecolor={blue}}

\usepackage{dcolumn}
\newcommand{\bra}[1]{\ensuremath{\langle #1|}}	
\newcommand{\ket}[1]{\ensuremath{|#1\rangle}}	
\newcommand{\abs}[1]{\ensuremath{\left |#1\right |}}
\renewcommand{\v}[1]{\ensuremath{\boldsymbol{#1}}}		
\newcommand{\vhat}[1]{\ensuremath{\hat{\boldsymbol{#1}}}}		
\renewcommand{\a}{\ensuremath{\alpha}} 
\newcommand{\w}{\ensuremath{\omega}} 
\def\h{\ensuremath{\hbar}} 
\newcommand{\g}{\ensuremath{\gamma}}

\begin{document}
\title{Search for a Variation of the Fine Structure around the Supermassive Black Hole in Our Galactic Center}

\author{A. Hees}
\email{aurelien.hees@obpsm.fr}
\affiliation{SYRTE, Observatoire de Paris, Universit\'e PSL, CNRS, Sorbonne Universit\'e, LNE, 61 avenue de l'Observatoire 75014 Paris, France}

\author{T. Do}
\affiliation{Department of Physics and Astronomy, University of California, Los Angeles, California 90095, USA}

\author{B.~M. Roberts}
\affiliation{SYRTE, Observatoire de Paris, Universit\'e PSL, CNRS, Sorbonne Universit\'e, LNE, 61 avenue de l'Observatoire 75014 Paris, France}
\affiliation{School of Mathematics and Physics, The University of Queensland, Brisbane, QLD 4072, Australia}

\author{A.~M.~Ghez}
\affiliation{Department of Physics and Astronomy, University of California, Los Angeles, California 90095, USA}

\author{S. Nishiyama}
\affiliation{Miyagi University of Education, 149 Aramaki-aza-aoba, Aoba-ku, Sendai, Miyagi 980-0845, Japan}

\author{R.~O.~Bentley}
\affiliation{Department of Physics and Astronomy, University of California, Los Angeles, California 90095, USA}

\author{A.~K.~Gautam}
\affiliation{Department of Physics and Astronomy, University of California, Los Angeles, California 90095, USA}

\author{S.~Jia}
\affiliation{Astronomy Department, University of California, Berkeley, CA 94720, USA}

\author{T.~Kara}
\affiliation{Miyagi University of Education, 149 Aramaki-aza-aoba, Aoba-ku, Sendai, Miyagi 980-0845, Japan}

\author{J.~R.~Lu}
\affiliation{Astronomy Department, University of California, Berkeley, CA 94720, USA}

\author{H.~Saida}
\affiliation{Daido University, 10-3 Takiharu-cho, Minami-ku, Nagoya, Aichi 457-8530, Japan}

\author{S.~Sakai}
\affiliation{Department of Physics and Astronomy, University of California, Los Angeles, California 90095, USA}

\author{M.~Takahashi}
\affiliation{Aichi University of Education, 1 Hirosawa, Igaya-cho, Kariya, Aichi 448-8542, Japan}

\author{Y.~Takamori}
\affiliation{National Institute of Technology, Wakayama College, 77 Noshima, Nada-cho, Gobo, Wakayama 644-0023, Japan}

\begin{abstract}
Searching for space-time variations of the constants of Nature is a promising way to search for new physics beyond General Relativity and the standard model motivated by unification theories and models of dark matter and dark energy. We propose a new way to search for a variation of the fine-structure constant using measurements of late-type evolved giant stars from the S-star cluster orbiting the supermassive black hole in our Galactic Center. A measurement of the difference between distinct absorption lines (with different sensitivity to the fine structure constant) from a star leads to a direct estimate of a variation of the fine structure constant between the star's location and Earth. Using spectroscopic measurements of 5 stars, we obtain a constraint on the relative variation of the fine structure constant below $10^{-5}$. This is the first time a varying constant of Nature is searched for around a black hole and in a high gravitational potential. This analysis shows new ways the monitoring of stars in the Galactic Center can be used to probe fundamental physics.
\end{abstract}

\maketitle
\clearpage
The current understanding of our Universe is based on the theory of General Relativity (GR) and on the Standard Model (SM) of particle physics. While both theories have been extremely successful, they are expected to break down at a certain point. In particular, a breaking of the Einstein equivalence principle is expected in various unification scenarios \cite{taylor:1988pi,*damour:1994fk,*damour:1994uq,fayet:2018aa,*fayet:2019aa}, in higher dimensional theories \cite{antoniadis:1998qd,*rubakov:2001it,*maartens:2010ek,*antoniadis:2011bl}, and by some models of dark matter \cite{arvanitaki:2015qy,*stadnik:2015yu,*hees:2018aa,carroll:2009fh} and dark energy \cite{khoury:2004uq,*khoury:2004fk,martins:2015qf,*martins:2015uq}. On a more philosophical note, the ``principle of absence of absolute structure'' led to many developments of extensions of physics where the constants of physics become dynamical entities, explicitly breaking the equivalence principle (see the discussion in section 2 of \cite{damour:2012zr}).

One way to test the equivalence principle is to search for space-time variations of the constants of Nature such as the fine structure constant $\alpha$, the mass of fermions and the quantum chromodynamics energy scale (see \cite{will:2014la} for a review of the tests of GR and \cite{uzan:2011vn} for a review of the search for varying constants). Various experiments using atomic clocks have provided stringent constraints on linear drifts of the constants of Nature at the level of $10^{-16}$ yr$^{-1}$ \cite{marion:2003zr,bize:2003ly,rosenband:2008fk,guena:2012ys,tobar:2013gf,leefer:2013xy,godun:2014sf,huntemann:2014nr}, on a dependency of the constants of Nature to the Sun gravitational potential at the level of $10^{-7}$ \cite{guena:2012ys,peil:2013ul,tobar:2013gf,leefer:2013xy,ashby:2018aa} or on harmonic variations of the constants of Nature \cite{van-tilburg:2015fj,*hees:2016uq}. A time variation of $\alpha$ has also been searched for using measurements of quasar absorption spectra \cite{webb:1999ao,*webb:2001fv,*murphy:2001rt,*webb:2003if,*murphy:2003dz,*srianand:2004wo,*chand:2004fr,*tzanavaris:2005hb,*tzanavaris:2007uk,*murphy:2008pd,*srianand:2008kx,*petitjean:2009aa,*webb:2011oj,*king:2012hb} providing constraint on $\Delta \alpha/\alpha$ at the level of $10^{-6}$ for cosmological redshift up to $z\sim 3$. A variation of $\alpha$ has also been constrained at the level of $10^{-3}$ for $z\sim10^3$ using cosmic microwave background measurements \cite{planck-collaboration:2014aa} and at a similar level for $z\sim 10^{10}$ by a big bang nucleosynthesis analysis \cite{avelino:2001ty}. Finally, a dependency of $\alpha$ on the gravitational potential has also been searched for using absorption lines from a white dwarf \cite{berengut:2013aa}. Although many searches for a variation of $\alpha$ have been performed, the question of its constancy around a black hole and around a supermassive body remains totally open.

The motion of the short-period stars (S-stars) orbiting around the $4\times 10^6 \, M_\odot$ supermassive black hole (SMBH) at the center of our Galaxy has been monitored for 25 years by two experiments: one carried out at the Keck Observatory~\cite{ghez:1998ve,*ghez:2000rt,*ghez:2003qv,*ghez:2005dq,*ghez:2005kx,*ghez:2008bs,*meyer:2012qf,*boehle:2016wu,*chu:2018aa,hees:2017aa,do:2019aa} and the other with the New Technology Telescope (NTT) and with the Very Large Telescope (VLT) \cite{genzel:1997zr,*eckart:1997ys,*schodel:2002bh,*eckart:2002qf,*eisenhauer:2003ty,*eisenhauer:2005dz,*gillessen:2009cr,*gillessen:2009jk,*gillessen:2017aa,*gravity:2019aa,gravity:2018aa,gravity:2019ab}. Recently, these measurements have opened a new window to probe fundamental physics around a SMBH. Measurements of the short-period star S0-2/S2  have been used to search for a fifth interaction~\cite{borka:2013nx,*zakharov:2016xy,*zakharov:2018aa,hees:2017aa}, to measure the relativistic redshift during its 2018 closest approach \cite{gravity:2018aa,do:2019aa} and to perform a null-redshift test \cite{gravity:2019ab}.

\begin{figure}[hbt]
\begin{center}
\includegraphics[width=0.35\textwidth]{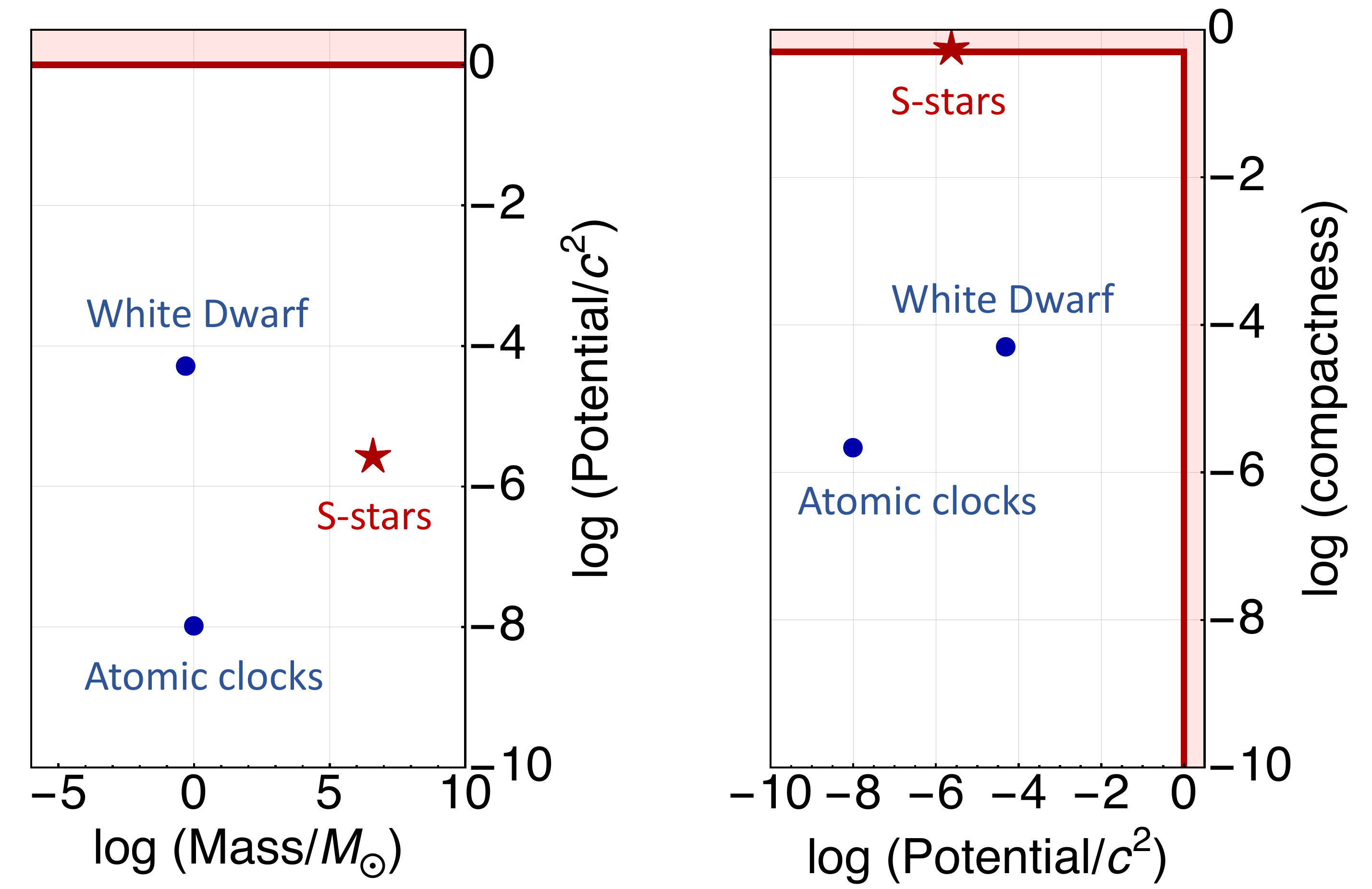}
\end{center}
\caption{Left: the gravitational potential probed by several searches for varying $\alpha$ against the mass of the central body that generates gravity in these tests. Right: the compactness $\Xi=GM/c^2R$ of the central body probed by several searches for varying $\alpha$ against the gravitational potential. Searches for varying $\alpha$ with S-stars explore a new region in this parameter space.}
\label{fig:gen}
\end{figure}

In this Letter, we present a novel search for variations of the fine structure constant around a SMBH using spectroscopic measurements from late-type evolved giant stars orbiting Sagittarius A$^*$ (Sgr A$^*$). This is the first search for a varying $\alpha$ around a compact object (the compactness of a celestial body can be characterized by $\Xi=GM/c^2R$ with $R$ the body's radius) and very massive object, exploring a new region in the parameter space (Fig.~\ref{fig:gen}). 
Probing the gravitational interaction in such a region is important to test for theories which exhibit screening mechanism in the solar system (see e.g.~\cite{khoury:2004uq,*khoury:2004fk,hinterbichler:2010fk,*hinterbichler:2011uq,vainshtein:1972ve,*deffayet:2002ly}) or for theories that exhibit a scalarization mechanism around a black hole~\cite{doneva:2018aa,*antoniou:2018aa,*silva:2018aa}.

Using spectroscopy, we can very precisely measure the wavelengths of atomic lines, which can be used to measure potential variations in the the fine structure constant. The spectroscopic observable is given by
\begin{equation}\label{eq:obs}
	\frac{\Delta \lambda_j}{\lambda_j}=\frac{\lambda_j(z,\alpha)-\lambda_j(z=0,\alpha_0)}{\lambda_j(z=0,\alpha_0)}=z-k_{\alpha,j}(1+z)\frac{\Delta \alpha}{\alpha}\, ,
\end{equation}
where $\lambda_j(z,\alpha)$ is the measured central wavelength of an absorption line $j$ at the GC for a value of $\alpha$ potentially different from the one on Earth $\alpha_0$ while $\lambda_j(0,\alpha_0)$ is the wavelength for the same line measured in the lab. In this expression, $z$ is the traditional Doppler shift (which includes the Newtonian velocity along the line of sight and the relativistic corrections on the redshift) and the last term encodes the impact due to a variation of $\alpha$ on the measurements. This term depends directly on $k_{\alpha,j}$, the sensitivity of the transition $j$ to $\alpha$ and defined by
\begin{equation}
	d\ln \nu_j=k_{\alpha,j}d\ln \alpha \, ,
\end{equation}
where $\nu_j=c/\lambda_j$ is the frequency of the absorption line $j$.

Late-type evolved stars are particularly useful for such an analysis because their cool atmospheres result in spectra that contain many strong atomic absorption lines which helps to maximize the number of different lines available for this analysis. Also importantly, their spectra contain lines that are related to atomic transitions with different sensitivity to $\alpha$ (the difference between the $k_\alpha$ coefficients are of the order of 1). In comparison, the more well studied young early-type S-stars such as S0-2 have atmospheres that are very hot, so their Near-Infrared (NIR) spectra contain only a few H and He absorption. 
Unfortunately, H and He lines have nearly the same sensitivity to $\alpha$ (the difference between the $k_\alpha$ coefficients are of the order of $10^{-5}$) making these lines insensitive to the effect of variations in $\alpha$. 

\begin{figure}[htb]
\begin{center}
\includegraphics[width=0.3\textwidth]{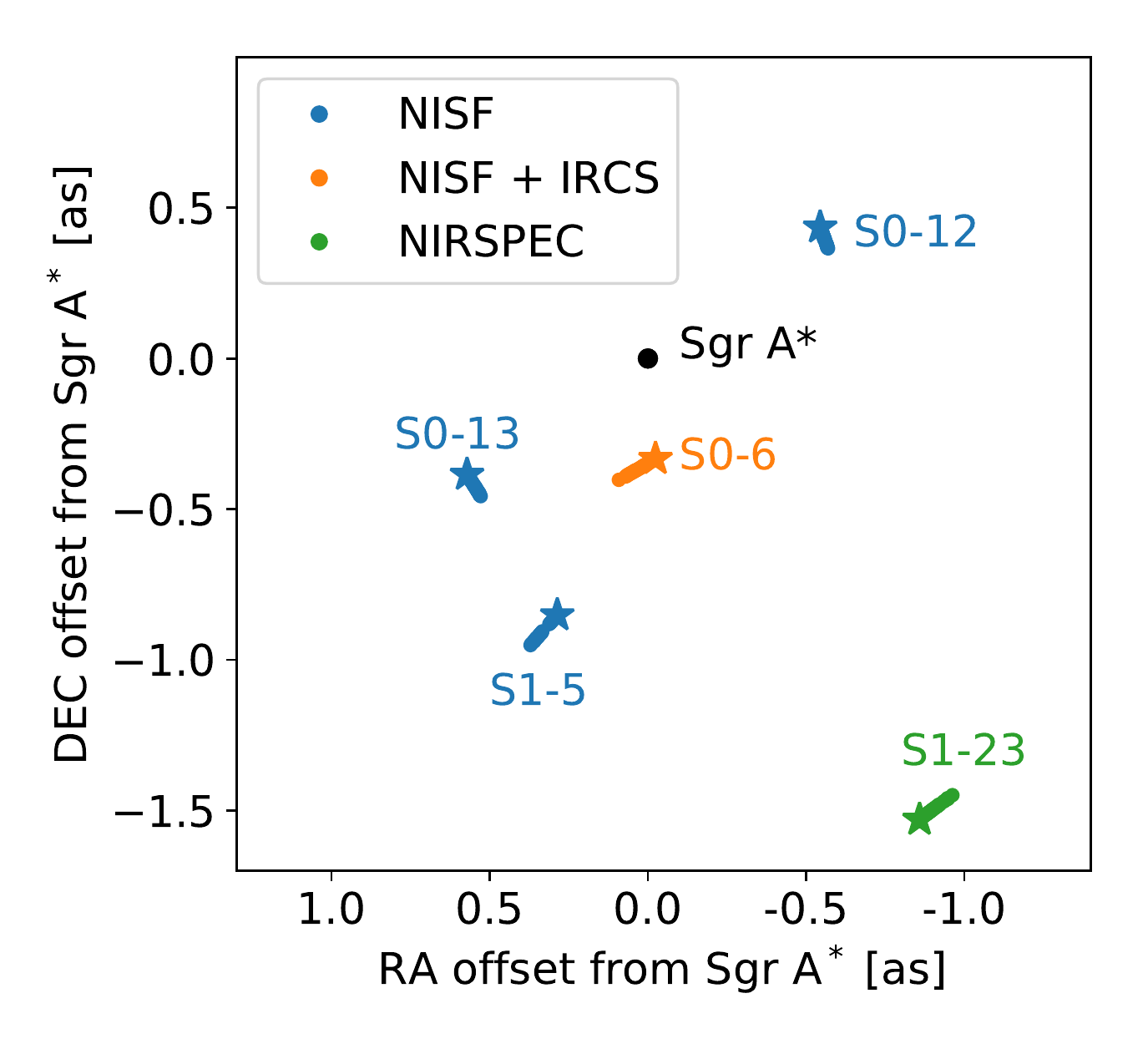}
\end{center}
\caption{Representation of the motion of the five stars considered in our analysis between 1995 and 2018. The large stars represent the positions of the stars in 2018. All these stars have a magnitude brighter than $M=15$. S0-6, S0-12, S0-13, S1-5 have been measured spectroscopically with the NIFS instrument with Gemini in 2018, S0-6 was also measured with the IRCS instrument from Subaru in 2018 and S1-23 with the NIRSPEC instrument at Keck in 2016.}
\label{fig:stars}
\end{figure}

In this work, we focus on 5 late-type giants in orbit around the supermassive black hole: S0-6, S0-12, S0-13, S1-5 and S1-23, see Fig.~\ref{fig:stars} and \cite{do:2013ty}. These are among the closest late-type stars to the black hole (0.4 to 1.5 arcseconds in projection) with NIR spectroscopic measurements. Observations of these stars include imaging data from Keck Observatory and spectroscopic data from Gemini North, Keck Observatory and Subaru telescope. The imaging data include both speckle imaging from the Near-Infrared Camera (NIRC) instrument (1995-2005) and adaptive optics data from the Near-Infrared Camera 2 (NIRC2) instrument (2005-2018). Details of the data reduction are given in \cite{do:2019aa}. 
NIR spectra were obtained with the Near-Infrared Integral Field Spectrometer (NIFS) instrument on Gemini North for the stars S0-6, S0-12, S0-13, and S1-5 using the K-band filter (0.95 to 2.40 $\mu$m) at a spectral resolution $R = \lambda/\Delta \lambda$ = 5000. We use spectra of these stars taken on May 13 and May 22, 2018. Additional spectroscopy of the star S0-6 was obtained using the Infrared Camera and Spectrograph (IRCS) instrument from Subaru. We use the the order 26 spectrum (2.16 to 2.22 $\mu m$) at a resolution of R = 20,000 (down-sampled to R = 15000) for this work. Details of the data reduction for IRCS is given in  \cite{Nishiyama2081} and \cite{Saida2019}.
Spectroscopy of the star S1-23 was obtained using the NIRSPEC instrument with laser-guide-star adaptive optics at Keck Observatory at a spectral resolution $R=$20,000. NIRSPEC was used in echelle mode with the K-filter. We use Orders 34 (2.24 to 2.27 $\mu$m) and 35 (2.18 to 2.21 $\mu$m) for this work. Details of the data reduction for the NIFS data and NIRSPEC data are given in \cite{do:2019aa} and \cite{do:2018}, respectively. 
All the NIFS, IRCS and NIRSPEC spectra have signal-to-noise ratios (SNR) greater than 20 in order to measure individual lines. 

\begin{figure*}[tbh]
\begin{center}
\includegraphics[width=1.\textwidth]{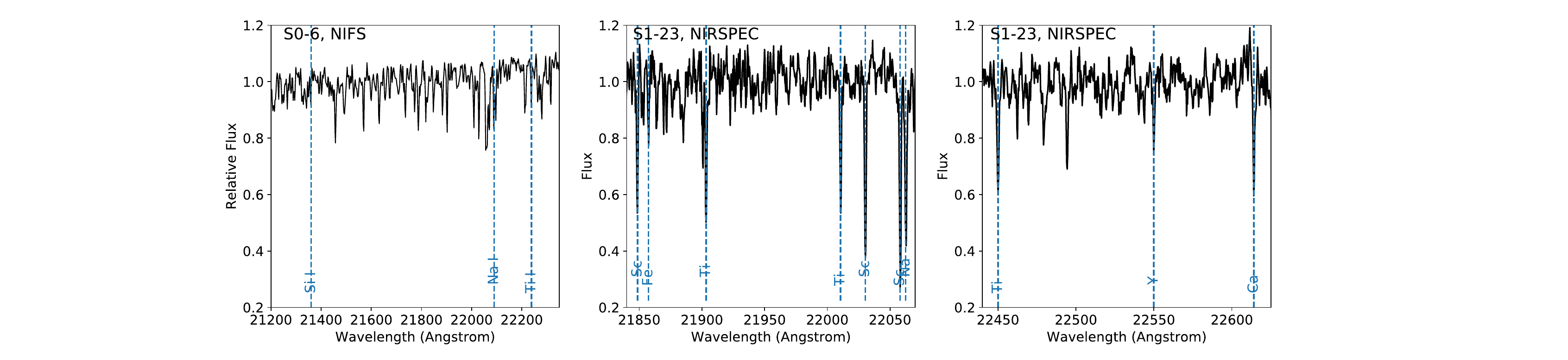}
\end{center}
\caption{Example spectra of the stars used in this work. \textbf{Left:} Spectrum of the star S0-6 observed using the NIFS spectrograph with a spectral resolution R = 5000. We have identified 3 lines that are suitable for the fine structure constant analysis. \textbf{Center \& Right:} Spectra of the star S1-23 observed with the NIRSPEC instrument with R = 20,000. This higher spectral resolution allows us to identify 10 atomic lines for use in this experiment. }
\label{fig:spectra}
\end{figure*}

We identified a number of strong and sufficiently isolated absorption lines for this experiment (Fig. \ref{fig:spectra}). In total, we measure 3 atomic lines from stars with NIFS spectra, 6 atomic lines with IRCS spectra, and 10 atomic lines from the NIRSPEC spectra (Tab.~\ref{tab:lines}). The higher spectral resolution from NIRSPEC compared to NIFS allows us to use more lines. 
We use a Gaussian fit to determine the central wavelength of each atomic transition and a Monte Carlo simulation to estimate the uncertainty. The fitted lines and uncertainties are reported in Sec.~II from the Supplemental Material \cite{supplemental}.

To determine the sensitivity coefficients $k_\alpha$ for each absorption line used in this study, we perform ab initio calculation of the spectrum for each element by solving the Dirac Hamiltonian in flat spacetime (curvature corrections are subdominant) and then compute $k_\alpha$ by a finite difference. The spectra are computed using a combination of the configuration interaction (CI) method with many-body perturbation theory (MBPT) \cite{dzuba:1996aa,*dzuba:1998aa,berengut:2016aa,dzuba:2017ab,geddes:2018aa} using the AMBiT software \cite{kahl:2019aa}. The full details of the calculations are presented in Appendix~\ref{ap:kalpha}; here we just outline briefly the main aspects of the method. The effective wavefunctions for a $M$-valence atom are expanded as a series of $M$-body Slater-determinants, where the expansion coefficients are found variationally by minimizing the energy (CI method). To greatly improve the accuracy, the effects of core-valence electron correlations are also included into the valence wavefunctions. This is done by modifying the effective Hamiltonian for the valence states using MBPT before the CI procedure is performed. Finally, to calculate the $k_\alpha$ coefficients, we explicitly make small variations into value of the fine structure constant ($\alpha \to \alpha \pm \delta\alpha $) in the code before the  equations are solved.

\begin{table}[h!]
\caption{\label{tab:lines}
Atomic properties of the absorption lines used in this analysis.  The wavelengths $\lambda$ are experimental values reported in \cite{NIST}. The sensitivity to the fine structure constant $k_\alpha$ is computed from ab initio calculation using the AMBIT software \cite{kahl:2019aa}, see the discussion in Appendix~\ref{ap:kalpha}. The last column indicates which instrument has been used to measured each line with: (a) NIFS spectrograph, (b) IRCS spectrograph, (c) NIRSPEC order34, (d) NIRSPEC order35. 
} 
\begin{ruledtabular}
\begin{tabular}{l ll ll r  l l} 
\multicolumn{1}{c}{}&	
\multicolumn{2}{c}{Lower}&
\multicolumn{2}{c}{Upper}&
\multicolumn{1}{c}{$\lambda$ [\AA]}&
\multicolumn{1}{c}{$k_\alpha$} &
\multicolumn{1}{c}{Inst.} \\
\hline
$_{14}$Si & $3s^23p4p$	& $^1D_{2}$	    & $3s^23p5s$	& $^1P^o_{1}$	& 21360.027	& 0.013(9)  & a \\
$_{11}$Na & $4s$	    & $^2S_{1/2}$	& $4p$	        & $^2P^o_{1/2}$	& 22089.728	& 0.004(2)  & a,b\\
$_{22}$Ti & $3d^34s$	& $^5P_{2}$	    & $3d^24s4p$	& $^5D^o_{2}$	& 22238.911	& -0.34(10) & a\\
\hline
$_{22}$Ti & $3d^34s$	& $^5P_{2}$   	& $3d^24s4p$	& $^5D^o_{1}$	& 22450.025	& -0.37(10) & c\\
$_{39}$Y  & $4d^25s$	& $^4F_{7/2}$	& $4d5s5p$	    & $^4F^o_{7/2}$	& 22549.938	& -0.88(6)  & c\\
$_{20}$Ca & $4s4d$  	& $^3D_{1}$	    & $4s4f$	    & $^3F^o_{2}$	& 22614.115	& -0.03(1)  & c\\
$_{21}$Sc & $3d^24s$	& $^4F_{3/2}$	& $3d4s4p$	    & $^2D^o_{3/2}$	& 21848.743	& -0.23(3)  & b,d\\
$_{39}$Fe &$3d^64s^2$	& $^3D_{3}$	    & $3d^64s4p$	& $^3P^o_{2}$	& 21857.345	& 0.56(28)  & d\\
$_{22}$Ti & $3d^34s$	& $^5P_{2}$	    & $3d^24s4p$	& $^5D^o_{3}$	& 21903.353	& -0.30(10) & b,d \\
$_{22}$Ti & $3d^34s$	& $^5P_{1}$     & $3d^24s4p$	& $^5D^o_{2}$	& 22010.501	& -0.31(9)  & b,d\\
$_{21}$Sc & $3d^24s$	& $^4F_{5/2}$	& $3d4s4p$   	& $^2D^o_{3/2}$	& 22030.179	& -0.25(4)  & b,d \\
$_{21}$Sc &  $3d^24s$	& $^4F_{9/2}$	& $3d4s4p$   	& $^4D^o_{7/2}$	& 22058.003	& -0.29(4)  & d\\
$_{11}$Na & $4s$	    & $^2S_{1/2}$   & $4p$	        & $^2P^o_{3/2}$	& 22062.485	& 0.007(2)  & b,d\\
\end{tabular}
\end{ruledtabular}
\end{table}

Bayesian inference is used to estimate $\Delta\alpha/\alpha$ from the measurements and Eq.~(\ref{eq:obs}). The measurement errors for each line are assumed to be independent and to be normally distributed such that the full likelihood is the product of the individual likelihoods
\begin{equation}
 \mathcal L_j \propto    \exp\left[ -\frac{1}{2\sigma_j^2}\left(\frac{\Delta\lambda_j}{\lambda_j}-z+k_{\alpha,j}(1+z)\frac{\Delta\alpha}{\alpha}\right)^2\right] \, ,
\end{equation}
where the subscript $j$ refers to a particular line. Uniform priors for $\Delta\alpha/\alpha$ and for $z$ are used and Gaussian priors are used for the sensitivity coefficients $k_{\alpha,j}$ with a mean value and an uncertainty quoted in Tab.~\ref{tab:lines}. We use the MULTINEST sampler \cite{feroz:2008qf,*feroz:2009xy} to sample the posterior probability distribution function. All the epochs are fitted simultaneously but offsets between the different instruments or filters are taken into account (this results in different fitted velocities for each instrument or filter).

No significant deviation of the fine structure constant is detected for any of the stars considered in this analysis. The posterior probability distributions for each star can be found in Sec.~IV from the supplemental material \cite{supplemental} and the 68\% confidence intervals are reported in Tab.~\ref{tab:dalpha}. The constraints derived from the NIRSPEC measurements are one order of magnitude better than from the NIFS instrument due to the better spectral resolution and more atomic lines observed with NIRSPEC.

\begin{table}[htb]
\caption{68\% confidence interval for $\Delta \alpha/\alpha$ and for $z$ estimated from different stars. An estimation of the gravitational potential $U$ at the location of the star is also provided (see Sec.~III from the supplemental material \cite{supplemental}). For S0-12, S0-13 and S1-5, two NIFS measurement epochs are combined with 3 absorption lines per epoch. S0-6 has been observed with NIFS and IRCS (each instrument has a different wavelength solution which reflects in an offset in their estimated $z$ value) providing 9 lines. S1-23 has been observed with two different filters (each filter has a different wavelength solution which reflects in an offset in their estimated $z$ value) providing 10 absorption lines.}
\label{tab:dalpha} 
\centering
\begin{tabular}{c c r l c }
\hline
Star     & $\frac{\Delta \alpha}{\alpha}$ & \multicolumn{2}{c}{$z.c$ [km/s]} &  $U/c^2$  \\
\hline
S0-6     & $(1.0 \pm 1.2)\times 10^{-4} $            & $71.0  $&$\pm 10.4$  & $2.4 \times 10^{-6}$ \\ 
         &                                           & $-3.6 $ & $\pm 5.3$\\
S0-12    & $(-0.3 \pm 1.4)\times 10^{-4} $ & $-57.3 $&$\pm 4.6$   & $1.6 \times 10^{-6}$ \\ 
S0-13    & $(\phantom{-} 0.03 \pm 3.5)\times 10^{-4} $ & $-61.6  $&$\pm 18.1$  & $9.4 \times 10^{-7}$ \\ 
S1-5     & $(- 0.7 \pm 2.4)\times 10^{-4} $ & $-3.7  $&$\pm 19.4$  & $6.5 \times 10^{-7}$ \\\hline  
S1-23    & $(\phantom{-} 0.9 \pm 5.8)\times 10^{-6} $ & $-311.4 $&$\pm 1.1$  & $4.6 \times 10^{-7}$ \\ 
         &                                           & $288.1   $&$\pm 0.7$  &  \\
\hline
\end{tabular}
\end{table}

A fit combining all the stars using a global likelihood (see Sec.~IV from the supplemental material \cite{supplemental} for more details)  provides a constraint of 
\begin{equation}
	\frac{\Delta \alpha}{\alpha}=\left(1.0\pm 5.8\right)\times 10^{-6}  \, ,
\end{equation} 
between the GC and Earth. This constraint is at the same level of magnitude as the ones obtained from quasar observations and is the first constraint on a possible variation of $\alpha$ around a BH.

In several alternative theories of gravitation, the fine structure constant becomes dependent on the gravitational potential (see e.g. \cite{bekenstein:1982zr,*sandvik:2002ly,*hees:2015ve}) and it is useful to consider the following parametrization
\begin{equation}
	\frac{\Delta \alpha}{\alpha}=\beta_\alpha \frac{\Delta U}{c^2}\, ,
\end{equation}
where $U$ is the Newtonian potential, $c$ the speed of light in a vacuum and where $\beta_\alpha$ depends on the fundamental parameters of the theory. An estimate of the gravitational potential probed by the 5 stars considered in this analysis is required in order to constrain the $\beta_\alpha$ parameter. We infer the radial acceleration experienced by the stars using 25 years of astrometric measurements of the GC. Between 1995 and 2005, speckle imaging data provides astrometric diffraction-limited measurements  of the central 5'' $\times$ 5'' of the GC \cite{ghez:1998ve,ghez:2000rt,ghez:2005dq,do:2019aa,chen:2019aa}. Between 2005 and 2018, adaptive optics (AO) imaging provides high-resolution images of the central 10''$\times$ 10'' of the GC \cite{ghez:2005kx,ghez:2008bs,boehle:2016wu,do:2019aa}. AO allows for more efficient observations at the diffraction limit, resulting in measurements typically one order of magnitude better than speckle observations. These astrometric measurements are aligned \cite{jia:2019aa} in a common reference frame defined by tying infrared observations of seven SiO masers~\cite{yelda:2010fj,*yelda:2013aa,sakai:2019ab} to their radio counterpart~\cite{reid:2007nr}. The resulting 2-D position measurements of the 5 stars considered in our analysis are provided in the Sec.~III from the supplemental material \cite{supplemental}. A polynomial fit of these measurements give an estimate of the 3D radial acceleration of these stars which is transformed into an estimate of the gravitational potential using the the SMBH mass $M=3.975 \times 10^6 M_\odot$ reported in \cite{do:2019aa}. The estimate of the gravitational potential experienced by each star is reported in Tab.~\ref{tab:dalpha}. A fit combining the measurements from the 5 stars and using the estimate from the gravitational potential from Tab.~\ref{tab:dalpha} leads to 
\begin{equation}
	\beta_\alpha=3.6\pm 12.0\, ,
\end{equation}
at 68\% confidence level. No deviation from GR is reported. This result is 8 orders of magnitude less constraining than a similar constraint obtained with atomic clocks around the Sun \cite{ashby:2018aa} and one order of magnitude less constraining than a similar result obtained around a white dwarf \cite{berengut:2013aa}. Nevertheless, this is the first time such a measurement is performed around a BH and around such a massive object (see Fig.~\ref{fig:gen}). Such a measurement is particularly appropriate to constrain the presence of a scalar field around a BH, which can naturally be enhanced in theories exhibiting BH scalarization mechanism, see e.g. \cite{doneva:2018aa,*antoniou:2018aa,*silva:2018aa,hees:2019ac}.

In conclusion, we propose a new method to search for a variation of the fine structure constant around a SMBH. Using existing measurements from NIFS spectrograph at GEMINI, from IRCS spectrograph from Subaru and from the NIRSPEC spectrograph at the Keck Observatory, we report a constraint on $\Delta\alpha/\alpha$ below $10^{-5}$ and a dependency of $\Delta\alpha/\alpha$ on the gravitational potential at the level of 10. This is the first time such a search has been performed around a compact object, around a BH or around such a massive object ($\sim 4\times 10^6 M_\odot$), see Fig.~\ref{fig:gen}. The results reported in this Letter could be improved by dedicated measurement sessions using high resolution spectrograph to measure spectrum of late-type stars closer to the GC. For example, a dedicated measurement of S0-6's spectrum using NIRSPEC could improve the limit on $\beta_\alpha$ by one order of magnitude. Furthermore, it is expected that future spectrographs with advanced AO systems such as KAPA at Keck \cite{kapa} would be able to measure spectrum of fainter late-type stars that are orbiting close to the GC (like e.g. S0-38/S38~\cite{boehle:2016wu}), allowing us to improve significantly the current constraints (by up to 4 orders of magnitude) and allowing one to probe the fine structure constant in a higher gravitational potential. Finally, after the search for a fifth force~\cite{borka:2013nx,*zakharov:2016xy,*zakharov:2018aa,hees:2017aa}, the relativistic redshift measurement of the star S0-2/S2~\cite{gravity:2018aa,do:2019aa} and a null test of the local position invariance using S0-2/S2 measurements~\cite{gravity:2019ab}, this analysis gives a new example of implications of GC measurements in term of fundamental physics.

\par {\it Acknowledgments.}

We thank the staff and astronomers at Keck Observatory and Gemini Observatory, especially Randy Campbell and Terry Stickel for all their help in obtaining the new NIRSPEC data. Some of the data presented herein were obtained at the W. M. Keck Observatory, which is operated as a scientific partnership among the California Institute of Technology, the University of California and the National Aeronautics and Space Administration and which was made possible by the generous financial support of the W. M. Keck Foundation. This work is partly based on observations obtained (Program ID GN-2018A-Q-123, PI: Do) at the Gemini Observatory, which is operated by the Association of Universities for Research in Astronomy, Inc., under a cooperative agreement with the NSF on behalf of the Gemini partnership: the National Science Foundation (United States), National Research Council (Canada), CONICYT (Chile), Ministerio de Ciencia, Tecnologia e Innovaci\'on Productiva (Argentina), Minist\'erio da Ciencia, Tecnologia e Inovacao (Brazil), and Korea Astronomy and Space Science Institute (Republic of Korea). This work is also based in part on data collected at Subaru Telescope, which is operated by the National Astronomical Observatory of Japan. S. N. was supported by JSPS KAKENHI, Grants No. JP15K13463, No. JP18K18760, and No. JP19H00695.  The authors wish to recognize that the summit of Maunakea has always held a very significant cultural role for the indigenous Hawaiian community. We are most fortunate to have the opportunity to observe from this mountain. Support for this work was provided by NSF AAG AST-1909554, the W.M. Keck Foundation, the Heising-Simons Foundation, and the Gordon and Betty Moore Foundation. Y. T. was supported by JSPS KAKENHI, Grant-in-Aid for Young Scientists (B) 26800150. We thank Julian Berengut for making the AMBiT software publicly available, and for discussions regarding its use.

\bibliography{GC_alpha}

\appendix

\section{Computation of the sensitivity of the transition energies to the fine structure constant}\label{ap:kalpha}

The atomic transition frequency can be expressed as
\begin{equation}
\omega = \omega_0 + qx,
\end{equation}
where $\omega_0$ is the nominal (observed laboratory) transition frequency,
$x$ parameterizes the variation in $\a$:
\begin{equation}
x\equiv (\alpha/\alpha_0)^2 - 1,
\end{equation}
and $q$ is the sensitivity factor.
Assuming the variation in $\a$ is small, $q$ can be determined from the derivative of $\omega$ to first-order around $x=0$.
In practice, this is done by calculating the relevant transition transition frequencies and explicitly allowing the value for $\a$ to vary inside the code (i.e., choosing small, non-zero values for $x$, denoted $\pm\delta x$).
Then, we have
\begin{equation}\label{eq:q-calc}
q = \frac{d\omega}{{d}x}\bigg|_{x=0} \approx \frac{\omega(+\delta x)-\omega(-\delta x)}{2\delta x}.
\end{equation}
We take values between $\delta x = 0.05$ and 0.1.
Multiple values are used to check for non-linearities.
Note that none of the atomic systems considered here have particularly high sensitivities (large $q$ values), so no non-linear effects are expected, and indeed none are observed. 
Much larger sensitivities can be found, e.g., in highly charged ions with large nuclear charges \citep{berengut:2012aa,safronova:2019aa}.
In general, $\delta x$ must be taken small enough so that the linear requirement of (\ref{eq:q-calc}) is valid, but must also be large enough for the variation $\delta\w$ to avoid numerical errors.

Note that $q$ has energy units.
It is convenient to further define a dimensionless sensitivity factor,
$k_\a$, defined via
\begin{equation}
\frac{\delta\w}{\w_0}=k_\a\frac{\delta\a}{\a},
\end{equation}
which is linked to $q$ as
\begin{equation}
k_\a = {2q}/{\w_0} \, .    
\end{equation}

For the infra-red transitions considered in this work, $\w_0$ is small compared to optical transitions from the ground state.
This manifests as an apparent enhancement in the magnitude of $k$; however, we note that this is partly compensated for by the uncertainty in the measurement of the observed frequencies (see also discussion in Ref.~\cite{kozlov:2018aa}).

\subsection{Method for calculation}

We use a method that is based on the combination of the configuration interaction (CI) method with many-body perturbation theory (MBPT) \cite{dzuba:1996aa,*dzuba:1998aa}.
The CI method accurately takes into account the valence--valence electron correlations, while the inclusion of MBPT allows an accurate treatment of the core--valence correlations.
The specific method we employ was developed 
in Ref.~\cite{kahl:2019aa}, see also Refs.~\cite{berengut:2016aa,dzuba:2017ab,geddes:2018aa}.
This implementation allows systems with a large number of valence electrons to be calculated accurately with reasonable computational resources.
Other CI methods for large valence systems have also proved successful \cite{dzuba:2019aa,lackenby:2019aa}.

The details of the method are presented in the above references, here we just give a brief overview.
The effective Hamiltonian for system of $M$ valence electrons is written as
\begin{equation}
 \hat H = \sum_i^M \hat h_1(\v{r}_i) + \sum_{i<j}^M\hat h_2(\v{r}_i, \v{r}_j),
\end{equation}
where $\hat h_1$ is the single-electron part of the Dirac Hamiltonian (in atomic units, $e=m_e=\h=1$),
\begin{equation}\label{eq:h1}
\hat h_1 = c\gamma^0\v{\g}\cdot\vhat{p} + c^2(\g^0-1) - V^{\rm nuc.} + V^{N-M} + \hat\Sigma_1,
\end{equation}
and $\hat h_2$ is the two-electron part,
\begin{equation} \label{eq:h2}
\hat h_2 = \frac{1}{\abs{\v{r}_i-\v{r}_j}}+ \hat\Sigma_2.
\end{equation}
In the above equations, $\v{\g}$ and $\g^0$ are Dirac matrices, $c=1/\a$ is the speed of light, $V^{\rm nuc.}\approx Z/r$ is the nuclear potential (formed assuming a Fermi-type nuclear distribution), and $V^{N-M}$ is the Hartree-Fock core potential, formed from the $N-M$ core electrons ($N$ is the total number of electrons).

The $\hat\Sigma$ terms are the correlation potentials, calculated to the second order in perturbation theory, without which the above equations would correspond to the conventional CI method.
The single electron correlation potential $\hat\Sigma_1$ represents the interaction of a single valence electron with the atomic core, and $\hat\Sigma_2$, a two-electron operator, represents the screening of the valence-valence Coulomb interaction by the core electrons (see Refs.~\cite{dzuba:1996aa,berengut:2006aa,dzuba:2017ab} for details).
Effective three-body MBPT contributions as introduced in Ref.~\cite{berengut:2008aa} are also included. For the systems considered here they make a reasonably small contribution, but do improve the accuracy.

The $M$-body CI wavefunctions are expanded as linear combinations of sets of $M$-body configuration functions (Slater determinants) that conserve a given $J_z$-parity symmetry,
\begin{equation}\label{eq:ci-exp}
\ket{\psi_{J_z^\pi}} = \sum_a c_a\ket{a,\, J_z^\pi}
\end{equation}
The $c$ expansion coefficients and CI energies are then found from the CI matrix eigenvalue problem
\begin{equation}
\sum_a\left(\bra{a}\hat H\ket{b} - E\delta_{ab}\right)c_a = 0.
\end{equation}
Many of the non-diagonal terms in the above matrix contribute negligibly.
In the ``emu CI'' approach \cite{kahl:2019aa}, some of these terms can be neglected from the calculations, which greatly reduces the number of integrals that need to be calculated and stored for the CI matrix.

The configuration functions are themselves formed from a set of single-particle basis orbitals, $\phi$.
These orbitals (which are also used to calculate the MBPT diagrams) are constructed using a potential that is formed by solving the Hartree-Fock equations self-consistently for a subset of the atomic orbitals.
In theory, with a complete enough basis, if all MBPT diagrams are calculated and the CI procedure converges, the choice of potential doesn't matter.
In practice, both the CI and MBPT parts are calculated approximately using a finite basis, and the choice of potential has a significant impact on the convergence of the calculations.
One common choice is to use the same $V^{N-M}$ potential as in Eq.~(\ref{eq:h1}), where the HF equations are solved for the closed-shell core electrons excluding all valence electrons.
Another is the $V^{N-1}$ potential, which corresponds to solving the HF equations including all but one of the electrons, and in general will involve partially-filled open shells.

The $V^{N-M}$ potential is convenient for evaluating the MBPT diagrams~\cite{dzuba:2005aa}, while
$V^{N-1}$ potential has the advantage of producing more realistic orbitals, allowing the CI expansion to converge more quickly.
Many of the states considered in this work lie relatively high in the spectrum.
In this regime, the core-valence correlation effects become less important, and the accuracy is mainly controlled the convergence of the CI part of the problem.
We use the $V^{N-M}$ potential for the two valence electron systems (Ca). 
For the other systems with more valence electrons, we use the $V^{N-1}$ potential which allows faster convergence of CI.
Calculations including $q$ factors for some of the lower-lying states of Ca were presented in Ref.~\cite{flambaum:2009ac}, and our calculations are in good agreement with those.

We treat silicon ($3s^23p^2,\,{}^3P_0$ ground state) as a two valence system, with the $3s^2$ orbitals remaining in the core.
We then allow single and double hole excitations into this $3s$ sub-shell, using  the particle-hole formalism described in Ref.~\cite{berengut:2016aa}.
Note that the sign of the $r_{ij}^{-1}$ term in Eq.~(\ref{eq:h2}) becomes negative for electron-hole interactions.
The affect of this is that some of the configuration functions are two-body, while others are four and six-body functions.
Since the configurations that include the $3s^2$ shell intact dominate the CI expansion, this method allows a significant reduction in the size of the CI matrix, while not neglecting the excitations of the $3s$ orbitals, which are nonetheless important.

For iron ($3d^64s^2\,{}^5D_4$ ground state), which has 8 valence electrons, we employ a restricted version of the method.
We use the $V^{N-1}$ potential, and allow only single excitations from specific reference configurations relevant to this work ($3d^64s^2$ for the even states, and $3d^64s4p$ for the odd states). 
We treat the $4s^2$ shell as though it were in the core, and allow single hole excitations into this shell.
Further, we do not include MBPT corrections (core--valence electron correlations).
This is a reasonable approximation, since the valence--valence electron correlations (treated at the CI level) dominate for systems with a large number of valence electrons.
Some calculations for iron, including for the $q$ sensitivities were presented in Ref.~\cite{dzuba:2008uq}, though not for the particular states of interest for this work.
We find that the $q$ factors are rather sensitive to configuration mixing; the same conclusion was found in Ref.~\cite{dzuba:2008uq}.
However, we find quite good agreement with experimental energies, and reasonable agreement between the $q$ values found in this work and those of Ref.~\cite{dzuba:2008uq} (our $q$ values are systematically around $\sim$\,20--30\% larger).
Conservatively, we assume a large 40\% uncertainty for all iron $q$ values, taking into account the spread of calculated $q$ values using different CI basis expansions.

We use a different approach for sodium, which has a single valence electron above a closed-shell core.
Namely, we employ the correlation potential method \cite{dzuba:1987aa}, in which the single-particle correlation potential is included into the Hartree-Fock equations
\begin{equation}
\left(\hat h^{\rm HF} + \hat \Sigma_1\right)\phi_a =\varepsilon_a \phi_a ,
\end{equation}
which are solved self-consistently for each of the relevant valence orbitals (which, when calculated this way, are known as Brueckner orbitals).
This is the same Hamiltonian as in Eq.~(\ref{eq:h1}); note that $V^{N-M}=V^{N-1}$ for $M=1$.
Again, the correlation potential is calculated to second-order in perturbation theory, though we note that it is possible to significantly improve the accuracy by using an all-order method \cite{dzuba:1989aa}.
Including $\Sigma$ into the self-consistent Hartree-Fock equations in this way effectively includes some classes of diagrams to all-orders~\cite{dzuba:1987aa}. 
Calculations of $q$-factors for the $3p$ and $4p$ states of sodium (referenced to the $3s$ ground state) were calculated in Ref.~\cite{berengut:2004aa}, using the same method employed here. 
In this work, we are interesting in transitions involving the $4s$ state.

\subsection{Uncertainty in q}

We estimate the uncertainty in each $q$ value by taking into account both the errors in the  energy-level determination, and the spread (stability) of the $q$ values as calculated with differing approximations (for example, by using a smaller basis for CI).
In most cases, the $q$ values for each level are found to be rather stable, and can be calculated with relatively high accuracy ($\sim1$--10\%).
The uncertainty in the $q$ values for the transitions of interest to this work, however, can be significantly larger.
This is because we consider transitions between states that are relatively high in the spectrum, and have similar $q$ values (leading to cancellation errors).

\subsection{Results}
Calculations of the energy levels, Land\'e $g$-factors, and $\delta\alpha$ sensitivity $q$-factors are shown for Y, Fe, Ti, Sc, Ca, Si, and Na in tables \ref{t:Y} through \ref{t:Na}.
The $q$ values are calculated with respect to the ground state.
The $g$-factors are useful for identifying states.
We don't present calculated $g$-factors for the lighter elements ($Z\leq20$), since they do not vary significantly from the non-relativistic values.
Table \ref{tab:qandk} shows the $q$ and $k$ sensitivities for the specific transitions of interest for this work.

\begin{table*}
\caption{\label{t:Y}
{\bf Yttrium ($Z=39$).}
Comparison of experimental and calculated excitation energies ($\Delta E \equiv E_{\rm Calc.} - E_{\rm Exp.}$),
with calculated Land\'e $g$-factors, and $\delta\alpha$ sensitivity $q$-factors (relative to the ground state).
Levels involved in transitions studied in this work are shown in bold.
}
\begin{ruledtabular}
\begin{tabular}{llrr D{.}{.}{0} rr D{(}{\,(}{-1}}
\multicolumn{1}{c}{Leading}&
\multicolumn{1}{c}{}&
\multicolumn{3}{c}{Energy (cm$^{-1}$)}&
\multicolumn{2}{c}{Land\'e $g$}&
\multicolumn{1}{c}{}\\
\cline{3-5}\cline{6-7}
\multicolumn{1}{c}{config.}&
\multicolumn{1}{c}{Term}&
\multicolumn{1}{c}{Exp.\cite{NIST}}&
\multicolumn{1}{c}{Calc.}&
\multicolumn{1}{c}{$\Delta E$}&
\multicolumn{1}{c}{Exp.\cite{NIST}}&
\multicolumn{1}{c}{Calc.}&
\multicolumn{1}{c}{$q$ (cm$^{-1}$)}\\
\hline
\multicolumn{2}{c}{Even Levels}&\multicolumn{5}{c}{}\\
$4d^25s$	& $^2D_{3/2}$	& 0	& 0	& 	& 0.80	& 0.80	&  \\ 
$4d^25s$	& $^2D_{5/2}$	& 530	& 581	& 50	& 1.20	& 1.20	& 530(70) \\ 
$4d^25s$	& $^4F_{3/2}$	& 10937	& 10778	& -159	& 0.40	& 0.40	& 2880(100) \\ 
$4d^25s$	& $^4F_{5/2}$	& 11079	& 10935	& -144	& 1.03	& 1.03	& 3030(110) \\ 
$4d^25s$	& \bf$^4F_{7/2}$	& \bf11278	& \bf11155	& \bf-123	& \bf1.24	& \bf1.24	& \bf3230(\bf120) \\ 
$4d^25s$	& $^4P_{1/2}$	& 15222	& 15426	& 204	& 2.61	& 2.66	& 2450(260) \\ 
$4d^25s$	& $^2F_{5/2}$	& 15327	& 15585	& 259	& 1.15	& 1.21	& 3740(580) \\ 
$4d^25s$	& $^4P_{3/2}$	& 15329	& 15544	& 215	& 1.62	& 1.72	& 2590(250) \\ 
$4d^25s$	& $^4P_{5/2}$	& 15477	& 15722	& 245	& 1.27	& 1.22	& 2230(240) \\ 
$4d^25s$	& $^2F_{7/2}$	& 15864	& 16176	& 312	& 1.17	& 1.14	& 3830(170) \\ 
$4d^25s$	& $^2D_{3/2}$	& 15994	& 16436	& 442	& 0.86	& 0.82	& 2890(180) \\ 
$4d^25s$	& $^2G_{7/2}$	& 18512	& 19176	& 664	& 0.90	& 0.89	& 3100(180) \\ 
$4d^25s$	& $^2P_{1/2}$	& 19238	& 19563	& 326	& 	& 0.67	& 3400(150) \\ 
$4d^25s$	& $^2S_{1/2}$	& 23465	& 23924	& 459	& 	& 2.00	& 1660(300) \\ 
$4d^3$	& $^4F_{7/2}$	& 29614	& 29468	& -146	& 	& 1.24	& 5490(260) \\ 
$5s26s$	& $^2S_{1/2}$	& 31672	& 31330	& -342	& 2.03	& 2.01	& -2720(170) \\ 
\multicolumn{2}{c}{Odd Levels}&\multicolumn{5}{c}{}\\
$5s^25p$	& $^2P^o_{1/2}$	& 10529	& 10174	& -356	& 0.63	& 0.67	& -2610(100) \\ 
$5s^25p$	& $^2P^o_{3/2}$	& 11360	& 11028	& -331	& 1.34	& 1.33	& -1710(50) \\ 
$4d5s5p$	& $^4F^o_{3/2}$	& 14949	& 14733	& -216	& 0.47	& 0.44	& 480(10) \\ 
$4d5s5p$	& $^4F^o_{5/2}$	& 15246	& 15044	& -202	& 1.08	& 1.05	& 770(20) \\ 
$4d5s5p$	& \bf$^4F^o_{7/2}$	& \bf15713	& \bf15556	& \bf-156	& \bf1.26	& \bf1.24	& \bf1280(\bf40) \\ 
$4d5s5p$	& $^2D^o_{5/2}$	& 16066	& 15835	& -231	& 1.20	& 1.19	& 1610(40) \\ 
$4d5s5p$	& $^2D^o_{3/2}$	& 16146	& 15921	& -226	& 0.80	& 0.79	& 1670(50) \\ 
$4d5s5p$	& $^4D^o_{1/2}$	& 16436	& 16155	& -281	& 0.01	& 0.00	& 1190(20) \\ 
$4d5s5p$	& $^4D^o_{3/2}$	& 16597	& 16334	& -263	& 1.22	& 1.17	& 1410(30) \\ 
$4d5s5p$	& $^4D^o_{5/2}$	& 16817	& 16567	& -249	& 1.38	& 1.37	& 1590(30) \\ 
$4d5s5p$	& $^4D^o_{7/2}$	& 17116	& 16888	& -228	& 1.42	& 1.43	& 1900(50) \\ 
$4d5s5p$	& $^4P^o_{1/2}$	& 18976	& 18815	& -161	& 	& 2.66	& 1390(30) \\ 
$4d5s5p$	& $^4P^o_{5/2}$	& 19148	& 19029	& -119	& 1.49	& 1.59	& 1600(50) \\ 
$4d5s5p$	& $^2F^o_{7/2}$	& 21915	& 21893	& -23	& 1.15	& 1.14	& 2450(60) \\ 
$4d5s5p$	& $^2P^o_{1/2}$	& 24699	& 24617	& -81	& 0.67	& 0.67	& 2260(70) \\ 
$4d5s5p$	& $^2F^o_{7/2}$	& 24900	& 24822	& -78	& 1.15	& 1.14	& 2150(120) \\ 
\end{tabular}
\end{ruledtabular}
\end{table*}

\begin{table*}
\caption{\label{t:Fe1}
{\bf Iron ($Z=22$), even states.}
Comparison of experimental and calculated excitation energies,
with calculated Land\'e $g$-factors, and $\delta\alpha$ sensitivity $q$-factors for some of the lower odd states of iron.
We estimate a $\sim$\,40\% uncertainty for the $q$ values.
} 
\begin{ruledtabular}
\begin{tabular}{lrrrrrrr}
\multicolumn{1}{c}{}&
\multicolumn{4}{c}{Energy (cm$^{-1}$)}&
\multicolumn{2}{c}{Land\'e $g$}&
\multicolumn{1}{c}{}\\
\cline{2-5}\cline{6-7}
\multicolumn{1}{c}{State}&
\multicolumn{1}{c}{Exp.\cite{NIST}}&
\multicolumn{1}{c}{Calc.}&
\multicolumn{1}{c}{$\Delta$}&
\multicolumn{1}{c}{Other \cite{dzuba:2008uq}}&
\multicolumn{1}{c}{Exp.\cite{NIST}}&
\multicolumn{1}{c}{Calc.}&
\multicolumn{1}{c}{$q$ (cm$^{-1}$)}
\\
\hline\\
\multicolumn{8}{l}{Leading configuration: $3d^64s^2_{}$ ($J=2$)}   \\
$^5D_{2}$	& 704	& 803	& -99	& 790	& 1.500	& 1.499	& 710	  \\
$^3P_{2}$	& 18378	& 19709	& -1331	& 	& 1.506	& 1.490	& 204	  \\
$^3F_{2}$	& 21039	& 23960	& -2921	& 	& 0.663	& 0.709	& 587	 \\
\multicolumn{8}{l}{Leading configuration: $3d^74s_{}$ ($J=2$)}   \\
$^5F_{2}$	& 7986	& 6592	& 1394	& 8078	& 1.000	& 1.000	& 2861	  \\
$^3F_{2}$	& 12969	& 14790	& -1821	& 14171	& 0.670	& 0.694	& 2960	  \\
$^5P_{2}$	& 17727	& 15979	& 1748	& 	& 1.820	& 1.813	& 2269	  \\
$^3P_{2}$	& 22838	& 23540	& -702	& 	& 1.498	& 1.354	& 1851	  \\
$^3P_{2}$	& 24336	& 25775	& -1440	& 	& 1.484	& 1.277	& 2609	  \\
$^3D_{2}$	& 26624	& 26836	& -212	& 	& 1.178	& 1.408	& 1677	  \\
$^1D_{2}$	& 28605	& 28513	& 92	& 	& 1.028	& 1.073	& 1957	  \\
$^3F_{2}$	& 36941	& 36385	& 555	& 	& 	& 0.687	& 1922	 \\
\multicolumn{8}{l}{Leading configuration: $3d^64s^2_{}$ ($J=3$)}   \\
$^5D_{3}$	& 416	& 429	& -13	& 464	& 1.500	& 1.499	& 396	  \\
$^3F_{3}$	& 20874	& 22709	& -1835	& 	& 1.073	& 1.061	& 246	  \\
$^3G_{3}$	& 24339	& 26316	& -1977	& 	& 0.761	& 0.783	& 1001	  \\
$\boldsymbol{^3D_{3}}$	&\bf 29372	&\bf 31928	&\bf -2556	&\bf 	&\bf 1.326	&\bf 1.327	&\bf 312	  \\
\multicolumn{8}{l}{Leading configuration: $3d^74s_{}$ ($J=3$)}   \\
$^5F_{3}$	& 7728	& 5404	& 2324	& 7779	& 1.250	& 1.247	& 2577	  \\
$^3F_{3}$	& 12561	& 12996	& -435	& 13702	& 1.086	& 1.072	& 2662	  \\
$^5P_{3}$	& 17550	& 16785	& 765	& 	& 1.666	& 1.654	& 2069	  \\
$^3G_{3}$	& 22249	& 20728	& 1522	& 	& 0.756	& 0.767	& 2498	  \\
$^3D_{3}$	& 26225	& 24808	& 1417	& 	& 1.335	& 1.332	& 1664	  \\
\multicolumn{8}{l}{Leading configuration: $3d^64s^2_{}$ ($J=4$)}   \\
$^5D_{4}$	& 	0& 	0& 	& 	& 1.500	& 1.499	& 	  \\
$^3H_{4}$	& 19788	& 19666	& 123	& 	& 0.811	& 0.828	& 442	  \\
$^3F_{4}$	& 20641	& 22425	& -1784	& 	& 1.235	& 1.144	& 438	  \\
$^3G_{4}$	& 24119	& 26316	& -2197	& 	& 1.048	& 1.053	& 1734	  \\
$^1G_{4}$	& 29799	& 31893	& -2094	& 	& 0.979	& 1.001	& 499	  \\
\multicolumn{8}{l}{Leading configuration: $3d^74s_{}$ ($J=4$)}   \\
$^5F_{4}$	& 7377	& 4377	& 3000	& 	& 1.350	& 1.346	& 2182	  \\
$^3F_{4}$	& 11976	& 12760	& -784	& 	& 1.254	& 1.247	& 2020	  \\
$^3G_{4}$	& 21999	& 19145	& 2855	& 	& 1.051	& 1.017	& 1951	  \\
$^1G_{4}$	& 24575	& 22943	& 1631	& 	& 1.001	& 0.978	& 2902	  \\
\end{tabular}
\end{ruledtabular}
\end{table*}

\begin{table*}
\caption{\label{t:Fe2}
{\bf Iron ($Z=22$), odd states.}
Comparison of experimental and calculated excitation energies,
with calculated Land\'e $g$-factors, and $\delta\alpha$ sensitivity $q$-factors for some of the lower odd states of iron.
We estimate a $\sim$\,40\% uncertainty for the $q$ values.
} 
\begin{ruledtabular}
\begin{tabular}{lrrrrrrrr}
\multicolumn{1}{c}{}&
\multicolumn{4}{c}{Energy (cm$^{-1}$)}&
\multicolumn{2}{c}{Land\'e $g$}&
\multicolumn{2}{c}{$q$ (cm$^{-1}$)}\\
\cline{2-5}\cline{6-7}\cline{8-9}
\multicolumn{1}{c}{State}&
\multicolumn{1}{c}{Exp.\cite{NIST}}&
\multicolumn{1}{c}{Calc.}&
\multicolumn{1}{c}{$\Delta$}&
\multicolumn{1}{c}{Other \cite{dzuba:2008uq}}&
\multicolumn{1}{c}{Exp.\cite{NIST}}&
\multicolumn{1}{c}{Calc.}&
\multicolumn{1}{c}{Calc.}&
\multicolumn{1}{c}{Other \cite{dzuba:2008uq}}\\
\hline\\
\multicolumn{9}{l}{Leading configuration: $3d^64s4p_{}$ ($J=1$)}   \\
$^7D^o_{1}$	& 20020	& 19855	& 165	& 19921	& 2.999	& 2.992	& 1568	& 1237 \\
$^7F^o_{1}$	& 23245	& 22711	& 534	& 22338	& 1.549	& 1.505	& 1530	& 1227 \\
$^5D^o_{1}$	& 26479	& 25364	& 1115	& 27094	& 1.495	& 1.480	& 2323	& 1616 \\
$^5F^o_{1}$	& 27666	& 26774	& 893	& 28213	& -0.012	& 0.023	& 2426	& 1680 \\
$^5P^o_{1}$	& 29733	& 29378	& 355	& 30118	& 2.487	& 2.494	& 1879	& 1594 \\
$^3D^o_{1}$	& 31937	& 31495	& 442	& 32750	& 0.513	& 0.559	& 2568	& 2119 \\
$^3P^o_{1}$	& 34363	& 34565	& -202	& 	& 1.496	& 1.496	& 2014	&  \\
$^5P^o_{1}$	& 37410	& 36116	& 1293	& 	& 2.502	& 2.493	& 903	&  \\
\multicolumn{9}{l}{Leading configuration: $3d^64s4p_{}$ ($J=2$)}   \\
$^7D^o_{2}$	& 19912	& 19568	& 345	& 19793	& 2.008	& 1.995	& 1445	& 1092 \\
$^7F^o_{2}$	& 23193	& 22628	& 564	& 22282	& 1.504	& 1.505	& 1504	& 1184 \\
$^7P^o_{2}$	& 24507	& 23859	& 648	& 23440	& 2.333	& 2.329	& 1539	& 1316 \\
$^5D^o_{2}$	& 26340	& 24955	& 1385	& 26924	& 1.503	& 1.480	& 2284	& 1450 \\
$^5F^o_{2}$	& 27560	& 26298	& 1262	& 28119	& 1.004	& 1.018	& 2421	& 1568 \\
$^5P^o_{2}$	& 29469	& 29019	& 450	& 29795	& 1.835	& 1.827	& 1620	& 1310 \\
$^3D^o_{2}$	& 31686	& 30817	& 869	& 32464	& 1.168	& 1.177	& 2647	& 1843 \\
$^3F^o_{2}$	& 32134	& 31492	& 642	& 33263	& 0.682	& 0.751	& 2624	& 2177 \\
$\boldsymbol{^3P^o_{2}}$	&\bf 33947	&\bf 34144	&\bf -197	&\bf 	&\bf 1.493	&\bf 1.494	&\bf 1596	&\bf  \\
\multicolumn{9}{l}{Leading configuration: $3d^64s4p_{}$ ($J=3$)}   \\
$^7D^o_{3}$	& 19757	& 19243	& 514	& 19611	& 1.746	& 1.747	& 1267	& 891 \\
$^7F^o_{3}$	& 23111	& 22468	& 643	& 22189	& 1.513	& 1.505	& 1436	& 1103 \\
$^7P^o_{3}$	& 24181	& 23466	& 715	& 23034	& 1.908	& 1.910	& 1235	& 983 \\
$^5D^o_{3}$	& 26140	& 24436	& 1704	& 26679	& 1.500	& 1.488	& 2122	& 1223 \\
$^5F^o_{3}$	& 27395	& 25626	& 1769	& 27947	& 1.250	& 1.259	& 2370	& 1402 \\
$^5P^o_{3}$	& 29056	& 28717	& 339	& 29340	& 1.657	& 1.661	& 1199	& 859 \\
$^3D^o_{3}$	& 31323	& 30457	& 865	& 32032	& 1.321	& 1.324	& 2196	& 1456 \\
$^3F^o_{3}$	& 31805	& 30787	& 1018	& 32883	& 1.086	& 1.071	& 2678	& 1808 \\
\multicolumn{9}{l}{Leading configuration: $3d^64s4p_{}$ ($J=4$)}   \\
$^7D^o_{4}$	& 19562	& 18933	& 630	& 19390	& 1.642	& 1.649	& 1081	& 662 \\
$^7F^o_{4}$	& 22997	& 22218	& 779	& 22062	& 1.493	& 1.507	& 1301	& 982 \\
$^7P^o_{4}$	& 23711	& 22986	& 726	& 22543	& 1.747	& 1.739	& 778	& 491 \\
$^5D^o_{4}$	& 25900	& 24081	& 1819	& 26428	& 1.502	& 1.500	& 1715	& 999 \\
$^5F^o_{4}$	& 27167	& 24999	& 2167	& 27702	& 1.355	& 1.350	& 2183	& 1180 \\
$^3F^o_{4}$	& 31307	& 29214	& 2093	& 32356	& 1.250	& 1.329	& 2342	& 1267 \\
\end{tabular}
\end{ruledtabular}
\end{table*}

\begin{table*}
\caption{\label{t:Ti}
{\bf Titanium ($Z=22$).}
Comparison of experimental and calculated excitation energies,
with calculated Land\'e $g$-factors, and $\delta\alpha$ sensitivity $q$-factors.
} 
\begin{ruledtabular}
\begin{tabular}{llrr D{.}{.}{0} rr D{(}{\,(}{-1}}
\multicolumn{1}{c}{Leading}&
\multicolumn{1}{c}{}&
\multicolumn{3}{c}{Energy (cm$^{-1}$)}&
\multicolumn{2}{c}{Land\'e $g$}&
\multicolumn{1}{c}{}\\
\cline{3-5}\cline{6-7}
\multicolumn{1}{c}{config.}&
\multicolumn{1}{c}{Term}&
\multicolumn{1}{c}{Exp.\cite{NIST}}&
\multicolumn{1}{c}{Calc.}&
\multicolumn{1}{c}{$\Delta E$}&
\multicolumn{1}{c}{Exp.\cite{NIST}}&
\multicolumn{1}{c}{Calc.}&
\multicolumn{1}{c}{$q$ (cm$^{-1}$)}\\
\hline
\multicolumn{2}{c}{Even Levels}&\multicolumn{5}{c}{}\\
$3d^24s2$	& $^3F_{2}$	& 0	& 0	& 	& 0.67	& 0.67	& 0\\
$3d^24s2$	& $^3F_{3}$	& 170	& 233	& 63	& 1.08	& 1.08	& 220(80)\\
$3d^34s$	& $^5F_{1}$	& 6557	& 7800	& 1243	& 0.00	& 0.00	& 1280(250)\\
$3d^34s$	& $^5F_{2}$	& 6599	& 7849	& 1250	& 1.00	& 0.99	& 1330(260)\\
$3d^34s$	& $^5F_{3}$	& 6661	& 7929	& 1268	& 1.25	& 1.25	& 1410(270)\\
$3d^24s2$	& $^1D_{2}$	& 7255	& 8486	& 1230	& 1.02	& 1.02	& 130(40) \\ 
$3d^24s2$	& $^3P_{0}$	& 8437	& 9264	& 828	& 	& 	& 90(40) \\ 
$3d^24s2$	& $^3P_{1}$	& 8492	& 9316	& 824	& 1.50	& 1.50	& 160(40) \\ 
$3d^24s2$	& $^3P_{2}$	& 8602	& 9509	& 906	& 1.48	& 1.49	& 350(60) \\ 
$3d^34s$	& $^3F_{2}$	& 11532	& 13812	& 2280	& 0.67	& 0.67	& 1520(310) \\ 
$3d^34s$	& $^3F_{3}$	& 11640	& 13963	& 2323	& 1.08	& 1.08	& 1650(340) \\ 
$3d^34s$	& \bf$^5P_{1}$	& \bf13982	& \bf16145	& \bf2163	& \bf2.50	& \bf2.50	& \bf1290(\bf210) \\ 
$3d^34s$	& \bf$^5P_{2}$	& \bf14028	& \bf16214	& \bf2186	& \bf1.83	& \bf1.82	& \bf1340(\bf220) \\ 
$3d^34s$	& $^5P_{3}$	& 14106	& 16328	& 2223	& 1.67	& 1.66	& 1440(240) \\ 
$3d^34s$	& $^3G_{3}$	& 15108	& 17901	& 2793	& 0.75	& 0.74	& 1350(220) \\ 
$3d^34s$	& $^3D_{1}$	& 17370	& 20683	& 3314	& 0.55	& 0.49	& 1180(250) \\ 
$3d^34s$	& $^3D_{3}$	& 17540	& 20925	& 3385	& 1.33	& 1.34	& 1430(240) \\ 
$3d^34s$	& $^3P_{0}$	& 17995	& 21261	& 3265	& 	& 	& 1220(250) \\ 
$3d^34s$	& $^3P_{1}$	& 18061	& 21391	& 3330	& 1.46	& 	& 1350(250) \\ 
$3d^34s$	& $^3P_{0}$	& 18818	& 21982	& 3164	& 	& 	& 1550(290) \\ 
$3d^34s$	& $^3P_{1}$	& 18826	& 21968	& 3142	& 1.49	& 	& 1510(280) \\ 
$3d^4$	& $^5D_{0}$	& 28773	& 32215	& 3442	& 	& 	& 580(1420) \\ 
$3d^4$	& $^3P_{0}$	& 34219	& 32300	& -1919	& 	& 	& 1430(1390) \\ 
$3d^24s4d$	& $^5D_{0}$	& 41871	& 38130	& -3742	& 	& 	& 1760(310) \\ 
\multicolumn{2}{c}{Odd Levels}&\multicolumn{5}{c}{}\\
$3d^24s4p$	& $^5G^o_{2}$	& 15877	& 15552	& -325	& 0.33	& 0.34	& 200(30) \\ 
$3d^24s4p$	& $^5G^o_{3}$	& 15976	& 15681	& -294	& 0.92	& 0.92	& 310(40) \\ 
$3d^24s4p$	& $^5F^o_{1}$	& 16817	& 16462	& -355	& 0.00	& 0.00	& 390(30) \\ 
$3d^24s4p$	& $^5F^o_{2}$	& 16875	& 16538	& -338	& 1.00	& 1.00	& 460(40) \\ 
$3d^24s4p$	& $^5F^o_{3}$	& 16961	& 16654	& -307	& 1.25	& 1.25	& 560(50) \\ 
$3d^24s4p$	& $^5D^o_{0}$	& 18463	& 18180	& 282	& 	&	& 00(40) \\ 
$3d^24s4p$	& \bf$^5D^o_{1}$	& \bf18483	& \bf18210	& \bf-273	& \bf1.50	& \bf1.50	& \bf520(\bf40) \\ 
$3d^24s4p$	& \bf$^5D^o_{2}$	& \bf18525	& \bf18275	& \bf-250	& \bf1.50	& \bf1.50	& \bf570(\bf40) \\ 
$3d^24s4p$	& \bf$^5D^o_{3}$	& \bf18594	& \bf18388	& \bf-206	& \bf1.50	& \bf1.50	& \bf660(\bf60) \\ 
$3d^24s4p$	& $^3F^o_{2}$	& 19323	& 19471	& 149	& 0.67	& 0.67	& 430(40) \\ 
$3d^24s4p$	& $^3F^o_{3}$	& 19422	& 19614	& 193	& 1.09	& 1.09	& 540(50) \\ 
$3d^24s4p$	& $^3D^o_{1}$	& 19938	& 20020	& 82	& 0.50	& 0.50	& 440(60) \\ 
$3d^24s4p$	& $^3D^o_{2}$	& 20006	& 20125	& 119	& 1.16	& 1.17	& 530(30) \\ 
$3d^24s4p$	& $^3D^o_{3}$	& 20126	& 20315	& 189	& 1.33	& 1.33	& 710(60) \\ 
$3d^24s4p$	& $^3G^o_{3}$	& 21469	& 22077	& 607	& 0.75	& 0.76	& 580(50) \\ 
$3d^24s4p$	& $^1D^o_{2}$	& 22081	& 22752	& 671	& 1.00	& 1.00	& 640(100) \\ 
$3d^24s4p$	& $^3S^o_{1}$	& 24921	& 25747	& 826	& 1.98	& 1.98	& 580(460) \\ 
$3d^34p$	& $^3D^o_{1}$	& 25318	& 26167	& 849	& 0.52	& 0.54	& 1000(270) \\ 
$3d^24s4p$	& $^3P^o_{1}$	& 25537	& 26337	& 800	& 1.50	& 1.48	& 490(220) \\ 
$3d^24s4p$	& $^3P^o_{0}$	& 25575	& 25649	& 74	& 	& 	& 640(40) \\ 
$3d^24s4p$	& $^5D^o_{0}$	& 25612	& 26303	& 690	& 	& 	& 450(200) \\ 
$3d^34p$	& $^5D^o_{0}$	& 29829	& 26589	& -3240	& 	& 	& 640(190) \\ 
$3d^24s4p$	& $^3P^o_{0}$	& 31686	& 30955	& -731	& 	& 	& 1740(160) \\ 
$3d^34p$	& $^3P^o_{0}$	& 33085	& 32811	& -274	& 	& 	& 320(350) \\ 
\end{tabular}
\end{ruledtabular}
\end{table*}

\begin{table*}
\caption{\label{t:Sc}
{\bf Scandium ($Z=21$).}
Comparison of experimental and calculated excitation energies,
with calculated Land\'e $g$-factors, and $\delta\alpha$ sensitivity $q$-factors.
} 
\begin{ruledtabular}
\begin{tabular}{llrr D{.}{.}{0} rr D{(}{\,(}{-1}}
\multicolumn{1}{c}{Leading}&
\multicolumn{1}{c}{}&
\multicolumn{3}{c}{Energy (cm$^{-1}$)}&
\multicolumn{2}{c}{Land\'e $g$}&
\multicolumn{1}{c}{}\\
\cline{3-5}\cline{6-7}
\multicolumn{1}{c}{config.}&
\multicolumn{1}{c}{Term}&
\multicolumn{1}{c}{Exp.\cite{NIST}}&
\multicolumn{1}{c}{Calc.}&
\multicolumn{1}{c}{$\Delta E$}&
\multicolumn{1}{c}{Exp.\cite{NIST}}&
\multicolumn{1}{c}{Calc.}&
\multicolumn{1}{c}{$q$ (cm$^{-1}$)}\\
\hline
\multicolumn{2}{c}{Even Levels}&\multicolumn{5}{c}{}\\
$3d^24s$	& $^2D_{3/2}$	& 0	& 0	& 	& 0.80	& 0.80	&  \\ 
$3d^24s$	& $^2D_{5/2}$	& 168	& 215	& 46	& 1.20	& 1.20	& 200(60) \\ 
$3d^24s$	& \bf$^4F_{3/2}$	& \bf11520	& \bf10919	& \bf-601	& \bf0.40	& \bf0.40	& \bf1050(70) \\ 
$3d^24s$	& \bf$^4F_{5/2}$	& \bf11558	& \bf10967	& \bf-590	& \bf1.03	& \bf1.03	& \bf1100(\bf70) \\ 
$3d^24s$	& $^4F_{7/2}$	& 11610	& 11036	& -574	& 1.24	& 1.24	& 1160(80) \\ 
$3d^24s$	& \bf$^4F_{9/2}$	& \bf11677	& \bf11130	& \bf-547	& \bf1.33	& \bf1.33	& \bf1250(\bf80) \\ 
$3d^24s$	& $^2F_{5/2}$	& 14926	& 14548	& -379	& 0.86	& 0.86	& 1180(60) \\ 
$3d^24s$	& $^2F_{7/2}$	& 15042	& 14703	& -339	& 1.13	& 1.14	& 1320(70) \\ 
$3d^24s$	& $^2D_{5/2}$	& 17013	& 17190	& 177	& 1.23	& 1.28	& 990(80) \\ 
$3d^24s$	& $^2D_{3/2}$	& 17025	& 17218	& 193	& 0.82	& 0.90	& 1030(70) \\ 
$3d^24s$	& $^4P_{1/2}$	& 17226	& 17308	& 82	& 2.66	& 2.67	& 950(90) \\ 
$3d^24s$	& $^4P_{3/2}$	& 17255	& 17346	& 91	& 1.72	& 1.63	& 1010(100) \\ 
$3d^24s$	& $^4P_{5/2}$	& 17307	& 17431	& 124	& 1.58	& 1.52	& 1100(60) \\ 
$3d^24s$	& $^2G_{9/2}$	& 20237	& 20540	& 303	& 1.10	& 1.11	& 1040(50) \\ 
$3d^24s$	& $^2G_{7/2}$	& 20240	& 20544	& 304	& 0.89	& 0.89	& 1040(60) \\ 
$3d^24s$	& $^2P_{1/2}$	& 20681	& 20891	& 209	& 0.67	& 0.67	& 1120(50) \\ 
$3d^24s$	& $^2P_{3/2}$	& 20720	& 20941	& 221	& 1.33	& 1.33	& 1170(50) \\ 
$3d^24s$	& $^2S_{1/2}$	& 26937	& 27719	& 782	& 	& 2.00	& 670(90) \\ 
$3d^3$	& $^4F_{7/2}$	& 33847	& 33588	& -258	& 1.23	& 1.24	& 1670(140) \\ 
$3d^3$	& $^4F_{9/2}$	& 33906	& 33672	& -234	& 1.33	& 1.33	& 1740(130) \\ 
$3d4s5s$	& $^4D_{1/2}$	& 34390	& 34015	& -375	& 0.00	& 0.00	& 220(0) \\ 
$3d4s5s$	& $^4D_{7/2}$	& 34567	& 34242	& -326	& 1.43	& 1.43	& 430(10) \\ 
$3d^3$	& $^4P_{1/2}$	& 36493	& 36511	& 19	& 2.63	& 2.65	& 1370(710) \\ 
$3d^3$	& $^2G_{9/2}$	& 37054	& 37204	& 150	& 1.11	& 1.11	& 1560(410) \\ 
$3d^3$	& $^2H_{9/2}$	& 39164	& 39369	& 205	& 	& 0.91	& 1660(1210) \\ 
\multicolumn{2}{c}{Odd Levels}&\multicolumn{5}{c}{}\\
$3d4s4p$	& $^4F^o_{3/2}$	& 15673	& 15334	& -338	& 0.43	& 0.44	& 160(10) \\ 
$3d4s4p$	& $^4F^o_{5/2}$	& 15757	& 15433	& -324	& 1.04	& 1.05	& 250(10) \\ 
$3d4s4p$	& $^4F^o_{7/2}$	& 15882	& 15585	& -297	& 1.23	& 1.24	& 400(20) \\ 
$3d4s4p$	& $^4D^o_{1/2}$	& 16010	& 15587	& -423	& 0.00	& 0.00	& 340(20) \\ 
$3d4s4p$	& $^4D^o_{3/2}$	& 16022	& 15617	& -405	& 1.04	& 1.13	& 380(40) \\ 
$3d4s4p$	& $^2D^o_{5/2}$	& 16023	& 15658	& -365	& 1.19	& 1.27	& 430(10) \\ 
$3d4s4p$	& $^4F^o_{9/2}$	& 16027	& 15744	& -283	& 1.33	& 1.33	& 570(20) \\ 
$3d4s4p$	& \bf$^2D^o_{3/2}$	& \bf16097	& \bf15772	& \bf-325	& \bf0.96	& \bf0.83	& \bf530(\bf40) \\ 
$3d4s4p$	& $^4D^o_{5/2}$	& 16141	& 15779	& -362	& 1.34	& 1.28	& 530(20) \\ 
$3d4s4p$	& \bf$^4D^o_{7/2}$	& \bf16211	& \bf15828	& \bf-383	& \bf1.43	& \bf1.43	& \bf580(\bf20) \\ 
$3d4s4p$	& $^4P^o_{1/2}$	& 18504	& 18206	& -298	& 2.53	& 2.64	& 390(40) \\ 
$3d4s4p$	& $^4P^o_{3/2}$	& 18516	& 18232	& -284	& 1.70	& 1.72	& 390(140) \\ 
$3d4s4p$	& $^4P^o_{5/2}$	& 18571	& 18303	& -269	& 1.60	& 1.60	& 490(10) \\ 
$4s^24p$	& $^2P^o_{1/2}$	& 18711	& 18743	& 32	& 0.78	& 0.69	& -390(280) \\ 
$4s^24p$	& $^2P^o_{3/2}$	& 18856	& 18889	& 34	& 1.36	& 1.34	& -200(230) \\ 
$3d4s4p$	& $^2F^o_{5/2}$	& 21033	& 20946	& -87	& 0.86	& 0.86	& 560(10) \\ 
$3d4s4p$	& $^2F^o_{7/2}$	& 21086	& 20995	& -91	& 1.14	& 1.14	& 610(10) \\ 
$3d4s4p$	& $^2P^o_{1/2}$	& 24657	& 24618	& -39	& 	& 0.67	& 420(90) \\ 
$3d4s4p$	& $^2F^o_{7/2}$	& 25725	& 25291	& -434	& 1.14	& 1.14	& 840(90) \\ 
$3d^24p$	& $^4G^o_{7/2}$	& 29096	& 28079	& -1017	& 0.98	& 0.98	& 1440(70) \\ 
$3d^24p$	& $^4G^o_{9/2}$	& 29190	& 28190	& -1000	& 1.16	& 1.17	& 1540(70) \\ 
$3d4s4p$	& $^2P^o_{1/2}$	& 30573	& 30601	& 28	& 0.68	& 0.67	& -190(180) \\ 
$3d^24p$	& $^4F^o_{9/2}$	& 31351	& 30297	& -1054	& 1.33	& 1.33	& 1680(80) \\ 
$3d^24p$	& $^2G^o_{9/2}$	& 33151	& 32290	& -862	& 1.06	& 1.11	& 1680(70) \\ 
$3d^24p$	& $^2H^o_{9/2}$	& 39153	& 39092	& -61	& 	& 1.04	& 1450(780) \\ 

\end{tabular}
\end{ruledtabular}
\end{table*}

\begin{table}[h!]
\caption{\label{t:Ca}
{\bf Calcium ($Z=20$).}
Comparison of experimental and calculated excitation energies, and calculated $q$-factors.
}
\begin{ruledtabular}
\begin{tabular}{llrr D{.}{.}{0}  D{(}{\,(}{-1}}
\multicolumn{1}{c}{Leading}&
\multicolumn{1}{c}{}&
\multicolumn{3}{c}{Energy (cm$^{-1}$)}&
\multicolumn{1}{c}{}\\
\cline{3-5}
\multicolumn{1}{c}{config.}&
\multicolumn{1}{c}{Term}&
\multicolumn{1}{c}{Exp.\cite{NIST}}&
\multicolumn{1}{c}{Calc.}&
\multicolumn{1}{c}{$\Delta E$}&
\multicolumn{1}{c}{$q$ (cm$^{-1}$)}\\
\hline
\multicolumn{2}{c}{Even Levels}&\multicolumn{3}{c}{}\\
$4s^2$	& $^1S_{0}$	& 0	& 0	& 	&0  \\ 
$3d4s$	& $^3D_{1}$	& 20335	& 20325	& -10	& 740(25) \\ 
$3d4s$	& $^3D_{2}$	& 20349	& 20350	& 1	& 760(26) \\ 
$3d4s$	& $^1D_{2}$	& 21850	& 22044	& 195	& 780(32) \\ 
$4s5s$	& $^3S_{1}$	& 31539	& 31704	& 164	& 130(2) \\ 
$4s5s$	& $^1S_{0}$	& 33317	& 33524	& 207	& 141(2) \\ 
$4s4d$	& $^1D_{2}$	& 37298	& 37506	& 208	& 370(22) \\ 
$4s4d$	& \bf$^3D_{1}$	& \bf37748	& \bf37932	& \bf184	& \bf243(\bf1) \\ 
$4s4d$	& $^3D_{2}$	& 37752	& 37937	& 185	& 247(2) \\ 
$4p^2$	& $^3P_{0}$	& 38418	& 38744	& 327	& 580(40) \\ 
$4p^2$	& $^3P_{1}$	& 38465	& 38795	& 330	& 640(39) \\ 
$4p^2$	& $^3P_{2}$	& 38552	& 38887	& 336	& 730(36) \\ 
$4s6s$	& $^3S_{1}$	& 40474	& 40604	& 129	& 150(2) \\ 
$4s6s$	& $^1S_{0}$	& 40690	& 40844	& 153	& 330(140) \\ 
$4p^2$	& $^1D_{2}$	& 40720	& 40919	& 199	& 690(56) \\ 
$4p^2$	& $^1S_{0}$	& 41786	& 42019	& 233	& 620(100) \\ 
$4s5d$	& $^3D_{1}$	& 42743	& 42919	& 176	& 204(7) \\ 
$4s5d$	& $^3D_{2}$	& 42745	& 42921	& 177	& 210(500) \\ 
$4s7s$	& $^3S_{1}$	& 43981	& 44235	& 255	& 156(3) \\ 
$4s7s$	& $^1S_{0}$	& 44277	& 44610	& 333	& 177(7) \\ 
$4s8s$	& $^1S_{0}$	& 45887	& 47669	& 1782	& 170(500) \\ 
\multicolumn{2}{c}{Odd Levels}&\multicolumn{3}{c}{}\\
$4s4p$	& $^3P^o_{0}$	& 15158	& 15398	& 240	& 132(3) \\ 
$4s4p$	& $^3P^o_{1}$	& 15210	& 15453	& 243	& 187(3) \\ 
$4s4p$	& $^3P^o_{2}$	& 15316	& 15565	& 249	& 298(5) \\ 
$4s4p$	& $^1P^o_{1}$	& 23652	& 23758	& 106	& 290(9) \\ 
$3d4p$	& $^3F^o_{2}$	& 35730	& 35891	& 161	& 1060(170) \\ 
$3d4p$	& $^1D^o_{2}$	& 35835	& 35780	& -56	& 980(330) \\ 
$4s5p$	& $^3P^o_{0}$	& 36548	& 36682	& 134	& 250(56) \\ 
$4s5p$	& $^3P^o_{1}$	& 36555	& 36690	& 136	& 260(55) \\ 
$4s5p$	& $^3P^o_{2}$	& 36575	& 36714	& 139	& 280(360) \\ 
$4s5p$	& $^1P^o_{1}$	& 36732	& 36911	& 180	& 320(29) \\ 
$3d4p$	& $^3D^o_{1}$	& 38192	& 38173	& -19	& 1100(19) \\ 
$3d4p$	& $^3D^o_{2}$	& 38219	& 38212	& -7	& 1130(20) \\ 
$3d4p$	& $^3P^o_{0}$	& 39333	& 39412	& 79	& 970(27) \\ 
$3d4p$	& $^3P^o_{1}$	& 39335	& 39422	& 87	& 980(27) \\ 
$3d4p$	& $^3P^o_{2}$	& 39340	& 39444	& 104	& 990(29) \\ 
$4s6p$	& $^1P^o_{1}$	& 41679	& 41847	& 168	& 410(94) \\ 
$4s4f$	& \bf$^3F^o_{2}$	& \bf42170	& \bf42430	& \bf260	& \bf170(\bf19) \\ 
$4s6p$	& $^3P^o_{0}$	& 42515	& 42670	& 155	& 180(14) \\ 
$4s7p$	& $^3P^o_{0}$	& 44956	& 45530	& 575	& 163(3) \\ 
$4s8p$	& $^3P^o_{0}$	& 46284	& 48852	& 2568	& 160(500) \\ 
\end{tabular}
\end{ruledtabular}
\end{table}

\begin{table}
\caption{\label{t:Si}
{\bf Silicon ($Z=14$).}
Comparison of experimental and calculated excitation energies, and calculated $q$-factors.
}
\begin{ruledtabular}
\begin{tabular}{llrr D{.}{.}{0}  D{(}{\,(}{-1}}
\multicolumn{1}{c}{Leading}&
\multicolumn{1}{c}{}&
\multicolumn{3}{c}{Energy (cm$^{-1}$)}&
\multicolumn{1}{c}{}\\
\cline{3-5}
\multicolumn{1}{c}{config.}&
\multicolumn{1}{c}{Term}&
\multicolumn{1}{c}{Exp.\cite{NIST}}&
\multicolumn{1}{c}{Calc.}&
\multicolumn{1}{c}{$\Delta E$}&
\multicolumn{1}{c}{$q$ (cm$^{-1}$)}\\
\hline
\multicolumn{2}{c}{Even Levels}&\multicolumn{3}{c}{}\\
$3s^23p^2$	& $^3P_{0}$	& 0	& 0	& 	&  \\ 
$3s^23p^2$	& $^3P_{1}$	& 77	& 18	& -59	& 85(4) \\ 
$3s^23p^2$	& $^3P_{2}$	& 223	& 186	& -37	& 240(9) \\ 
$3s^23p^2$	& $^1D_{2}$	& 6299	& 6456	& 157	& 160(10) \\ 
$3s^23p^2$	& $^1S_{0}$	& 15394	& 15640	& 245	& 170(10) \\ 
$3s^23p4p$	& $^1P_{1}$	& 47284	& 47024	& -260	& 39(5) \\ 
$3s^23p4p$	& $^3D_{1}$	& 48020	& 47738	& -282	& -62(6) \\ 
$3s^23p4p$	& $^3D_{2}$	& 48102	& 47827	& -275	& 2(4) \\ 
$3s^23p4p$	& $^3P_{0}$	& 49028	& 48784	& -245	& -42(3) \\ 
$3s^23p4p$	& $^3P_{1}$	& 49061	& 48803	& -257	& -21(13) \\ 
$3s^23p4p$	& $^3P_{2}$	& 49189	& 48946	& -243	& 140(9) \\ 
$3s^23p4p$	& $^3S_{1}$	& 49400	& 49112	& -288	& 146(5) \\ 
$3s^23p4p$	& \bf$^1D_{2}$	& \bf50189	& \bf49997	& \bf-192	& \bf100(\bf11) \\ 
$3s^23p4p$	& $^1S_{0}$	& 51612	& 51548	& -64	& 95(10) \\ 
$3s^23p5p$	& $^1P_{1}$	& 56780	& 58674	& 1894	& -33(18) \\ 
$3s^23p5p$	& $^3D_{2}$	& 57017	& 58929	& 1912	& -31(12) \\ 
$3s^23p5p$	& $^3P_{0}$	& 57296	& 59225	& 1930	& -53(10) \\ 
$3s^23p5p$	& $^1S_{0}$	& 58312	& 60212	& 1901	& 120(20) \\ 
\multicolumn{2}{c}{Odd Levels}&\multicolumn{3}{c}{}\\
$3s^3p3$	& $^5S^o_{2}$	& 33326	& 33348	& 22	& 580(20) \\ 
$3s^23p4s$	& $^3P^o_{0}$	& 39683	& 39418	& -265	& -157(3) \\ 
$3s^23p4s$	& $^3P^o_{1}$	& 39760	& 39504	& -257	& -90(3) \\ 
$3s^23p4s$	& $^3P^o_{2}$	& 39955	& 39707	& -248	& 130(10) \\ 
$3s^23p4s$	& $^1P^o_{1}$	& 40992	& 40764	& -227	& 69(13) \\ 
$3s^3p3$	& $^3D^o_{1}$	& 45276	& 45715	& 438	& 330(20) \\ 
$3s^3p3$	& $^3D^o_{2}$	& 45294	& 45687	& 393	& 350(20) \\ 
$3s^23p3d$	& $^1D^o_{2}$	& 47352	& 47552	& 201	& 140(10) \\ 
$3s^23p3d$	& $^3F^o_{2}$	& 49851	& 49930	& 79	& -24(9) \\ 
$3s^23p3d$	& $^3P^o_{2}$	& 50500	& 50558	& 58	& 150(15) \\ 
$3s^23p3d$	& $^3P^o_{1}$	& 50566	& 50673	& 107	& 220(15) \\ 
$3s^23p3d$	& $^3P^o_{0}$	& 50602	& 50679	& 77	& 260(20) \\ 
$3s^23p3d$	& $^1P^o_{1}$	& 53387	& 53213	& -174	& 56(12) \\ 
$3s^23p3d$	& $^3D^o_{1}$	& 54185	& 54056	& -130	& -86(5) \\ 
$3s^23p5s$	& $^3P^o_{0}$	& 54245	& 53982	& -263	& -139(3) \\ 
$3s^23p5s$	& $^3P^o_{1}$	& 54314	& 54344	& 30	& 150(20) \\ 
$3s^23p5s$	& \bf$^1P^o_{1}$	& \bf54871	& \bf54665	& \bf-206	& \bf130(\bf20) \\ 
$3s^23p4d$	& $^3P^o_{0}$	& 56733	& 57702	& 969	& 310(60) \\ 
$3s^23p6s$	& $^3P^o_{0}$	& 59221	& 61135	& 1914	& -120(10) \\ 
$3s^23p4d$	& $^3P^o_{0}$	& 60043	& 61752	& 1709	& 230(10) \\ 
\end{tabular}
\end{ruledtabular}
\end{table}

\begin{table}[h!]
\caption{\label{t:Na}
{\bf Sodium ($Z=11$).}
Comparison of experimental and calculated excitation energies, and calculated $q$-factors.}
\begin{ruledtabular}
\begin{tabular}{llrr D{.}{.}{0}  D{(}{\,(}{-1}}
\multicolumn{1}{c}{}&
\multicolumn{1}{c}{}&
\multicolumn{3}{c}{Energy (cm$^{-1}$)}&
\multicolumn{1}{c}{}\\
\cline{3-5}
%
\multicolumn{2}{c}{State}&
\multicolumn{1}{c}{Exp. \cite{NIST}}&
\multicolumn{1}{c}{Calc.}&
\multicolumn{1}{c}{$\Delta E$}&
\multicolumn{1}{c}{$q$ (cm$^{-1}$)}\\
\hline
\multicolumn{2}{c}{Even Levels}&\multicolumn{3}{c}{}\\
$3s$	& $^2S_{1/2}$	& 0	& 0	&   &  0	\\
$\boldsymbol{4s}$	& $\boldsymbol{^2S_{1/2}}$	& \bf25740	& \bf25661	& \bf-79	& \bf45(\bf3)	\\ 
\multicolumn{2}{c}{Odd Levels}&\multicolumn{3}{c}{}\\
$3p$	& $^2P^o_{1/2}$	& 16956	& 16860	& -96	& 45(3)	\\ 
$3p$	& $^2P^o_{3/2}$	& 16973	& 16878	& -95	& 64(4)	\\ 
$\boldsymbol{4p}$	& $\boldsymbol{^2P^o_{1/2}}$	& \bf30267	& \bf30161	& \bf-106	& \bf54(\bf3)	\\ 
$\boldsymbol{4p}$	& $\boldsymbol{^2P^o_{3/2}}$	& \bf30273	& \bf30167	& \bf-106	& \bf60(\bf4)	\\ 
\end{tabular}
\end{ruledtabular}
\end{table}

\begin{table}[h!]
\caption{\label{tab:qandk}
Transition frequencies, and calculated $q$ and $k$ sensitivity coefficients.
} 
\begin{ruledtabular}
\begin{tabular}{ll ll r D{(}{(}{-1} r  D{(}{(}{5}}
\multicolumn{4}{c}{}&
\multicolumn{2}{c}{(cm$^{-1}$)}&
\multicolumn{1}{c}{(\AA)}&
\multicolumn{1}{c}{}\\
\cline{5-6}
\multicolumn{2}{c}{Lower}&
\multicolumn{2}{c}{Upper}&
\multicolumn{1}{c}{$\omega$ \cite{NIST}}&

\multicolumn{1}{c}{$q$}&
\multicolumn{1}{c}{$\lambda$}&
\multicolumn{1}{c}{$k_\alpha$}\\
\hline
\\\multicolumn{7}{l}{\bf Y ($Z=39$)} \\
$4d^25s$	& $^4F_{7/2}$	& $4d5s5p$	& $^4F^o_{7/2}$	& 4434.6 	& -1950(130) & 22549.938 & -0.88(6) \\
\\\multicolumn{7}{l}{\bf Fe ($Z=26$)}\\
$3d^64s^2$	& $^3D_{3}$	& $3d^64s4p$	& $^3P^o_{2}$	& 4575.1	& 1284(650)	& 21857.345 & 0.56(28) \\
\\\multicolumn{7}{l}{\bf Ti ($Z=22$)} \\
$3d^34s$	& $^5P_{1}$	& $3d^24s4p$	& $^5D^o_{2}$	& 4543.3	& -720(210)	& 22010.501 & -0.31(9) \\
$3d^34s$	& $^5P_{2}$	& $3d^24s4p$	& $^5D^o_{1}$	& 4454.3	& -820(220)	& 22450.025 & -0.37(10) \\
            & 			& $3d^24s4p$	& $^5D^o_{2}$	& 4496.6	& -770(220)	& 22238.911 & -0.34(10) \\
            &        	& $3d^24s4p$	& $^5D^o_{3}$	& 4565.5	& -680(230)	& 21903.353 & -0.30(10) \\			
\\\multicolumn{7}{l}{\bf Sc ($Z=21$)} \\
$3d^24s$	& $^4F_{3/2}$	& $3d4s4p$	& $^2D^o_{3/2}$	& 4576.9	& -520(80)	& 21848.743 & -0.23(3) \\
$3d^24s$	& $^4F_{5/2}$	& $3d4s4p$	& $^2D^o_{3/2}$	& 4539.2	& -570(80)	& 22030.179 & -0.25(4) \\
$3d^24s$	& $^4F_{9/2}$	& $3d4s4p$	& $^4D^o_{7/2}$	& 4533.5	& -670(80)	& 22058.003 & -0.29(4) \\
\\\multicolumn{7}{l}{\bf Ca ($Z=20$)} \\
$4s4d$	& $^3D_{1}$	& $4s4f$	& $^3F^o_{2}$	& 4422.0	& -73(19)	& 22614.115 & -0.03(1) \\
\\\multicolumn{7}{l}{\bf Si ($Z=14$)} \\
$3s^23p4p$	& $^1D_{2}$	& $3s^23p5s$	& $^1P^o_{1}$	& 4681.6	& 30(21)	& 21360.027 & 0.013(9) \\
\\\multicolumn{7}{l}{\bf Na ($Z=11$)} \\
$4s$	& $^2S_{1/2}$	& $4p$	& $^2P^o_{1/2}$	& 4527.0	& 9(5)	& 22089.728 & 0.004(2) \\
	& 	& $4p$	& $^2P^o_{3/2}$	& 4532.6	& 15(5)	& 22062.485 & 0.007(2) \\
\end{tabular}
\end{ruledtabular}
\end{table}
\end{document}